\def\to{\hbox{$\,$--$\,$}}
\def\minus{\hbox{$\,$--$\,$}}
\def\muspc{\hskip 0.15 em}

\def\mag{\hbox{$\;.\!\!\!^m$}}
\input psfig
\documentclass{aa-osservatorio-b}
\usepackage{graphics}
\begin{document}

\thesaurus{06(08.03.1, 08.16.4, 08.05.3,
Sagittarius dwarf; 11.12.1)}

\title{Are the Bulge C{\to}stars related to the 
Sagittarius dwarf galaxy?}

\subtitle{II. Metallicity -- a link with the Galactic disc  
}

\author{Yuen K. Ng\thanks{\emph{Present address (per 1 October):}
Space Research Organisation Netherlands
(SRON), Sorbonnelaan 2, 3584~CA \ Utrecht, the Netherlands}}

\institute{Padova Astronomical Observatory,
           Vicolo dell'Osservatorio 5, I-35122 Padua, Italy 
           ({\tt yuen\char64pd.astro.it})}
\date{Received 20 April 1998 / Accepted 23 July 1998}

\maketitle

\markboth{Yuen K. Ng:\ Are the Bulge carbon stars related to the
Sagittarius dwarf galaxy?}
{Yuen K. Ng:\ Are the Bulge carbon stars related to the
Sagittarius dwarf galaxy?}

\begin{abstract}
The photometric estimate of the
metallicity and the age of the Azzopardi et~al.\ (\cite{ALRW91}) 
carbon stars (Ng \cite{Ng97}) 
is revised to respectively \mbox{Z\muspc$\simeq$\muspc0.004}
and \mbox{$\sim$\muspc0.1~Gyr}.
Under the hypothesis that the carbon stars are located at a 
distance related to the Sagittarius dwarf galaxy,
the broad velocity dispersion of the stars can only be explained
if they were formed out of Galactic material 
during a recent crossing of the Sagittarius
dwarf galaxy through the Galactic plane. 
\keywords{stars: carbon; evolution 
--- galaxies: individual: Sagittarius dwarf -- Local Group}
\end{abstract}

\section{Introduction}
\label{Introduction}
The near-IR colours and medium-low resolution spectra 
(Azzopardi et~al.\ \cite{ALRW91} -- hereafter referred to as ALRW91, 
Tyson{\muspc\&\muspc}Rich \cite{TR91} --
hereafter referred to as TR91, Westerlund et~al.\ \cite{Westerlund91ea})
obtained for the so-called `bulge' carbon stars, 
identified by Azzopardi et~al.\ (\cite{ALR85}, \cite{ALR88}),
show similarities with the low- to medium bolometric
luminosity SMC carbon stars. 
The main difference is that the galactic carbon 
stars are photometrically bluer and that they 
have spectroscopically stronger 
NaD-doublets.
\par
The radial velocities together with the direction in which
these stars are located 
suggest a Bulge membership.
In addition,
a high metallicity for the Bulge lead 
TR91 and Westerlund et~al.\ (\cite{Westerlund91ea}) to suggest
that the stars should be old
and posses a mass of about 0.8~M$_\odot$,
while evolutionary calculations (Boothroyd et~al.\ \cite{Boothroyd93ea}, 
Groenewegen{\muspc\&\muspc}de~Jong \cite{GdJ93}, 
Groenewegen et~al.\ \cite{Groenewegen95ea}, 
Marigo \cite{Marigo98},
Marigo et~al.\ 1996ab) demonstrate that the initial mass 
of carbon stars in general has to be at least 
\mbox{$\sim$1.2~M$_\odot$} ($t\!\la\!4$~Gyr)
for both \mbox{Z\muspc=\muspc0.004} and \mbox{Z\muspc=\muspc0.008}. 
Furthermore, the initial mass increases 
towards higher metallicity 
(Lattanzio \cite{Lattanzio89}).
The ALRW91 C-stars are a mystery 
(Lequeux \cite{Lequeux90}, TR91, 
Westerlund et~al.\ \cite{Westerlund91ea}, 
Chiosi et~al.\ \cite{CBB92},
Azzopardi \cite{Azzopardi94}), because they 
are in bolometric luminosity about 2\mag5 too faint
to be regarded as genuine AGB (Asymptotic Giant Branch) stars, 
if located in the metal-rich Bulge.
\par
The serendipitous identification of the 
Sagittarius dwarf galaxy (SDG) was made by 
Ibata et~al.\ (\cite{IGI94}, \cite{IGI95}).
The $\sim$2\mag5 difference of the distance modulus 
between the dwarf galaxy and the Galactic Centre at 8~kpc
lead Ng\mbox{\muspc\&\muspc}Schultheis (\cite{NS97}, hereafter referred
to as NS97) to suggest that the ALRW91 C-stars
could actually be located at the distance of the 
dwarf galaxy. Its presence
was unknown at the time when the C-stars were identified
and a different location 
could solve the standing question about the origin 
of the `bulge' carbon stars.
\par
Ng (\cite{Ng97}, \cite{Ng98}) analysed the possibility 
that the ALRW91 C-stars are related to the SDG.
With this hypothesis there is no need exotic 
stellar evolutionary scenarios to explain these stars. 
Ng demonstrated that
the photometric sequence of the ALRW91 C-stars 
is not exceptional, but comparable with the 
sequence found for the SMC. 
The estimated metallicity and age were respectively 
\mbox{Z\muspc$\simeq$\muspc0.008}
and \mbox{0.1\to1~Gyr}.
\par
The organisation of the paper is that Sect.~\ref{SDG}
begins with an overview about the age and metallicities 
of the various populations identified in the SDG.
Sect.~\ref{Isochrones} deals with the improvement of
the photometric metallicity and age estimates of the ALRW91 C-stars. 
In Sect.~\ref{Vdispersion} additional constraints are obtained 
from the velocity dispersion.
In Sect.~\ref{Motion} 
it is further argued that the present position
of the C-stars does not violate any of 
the reliable observational constraints.
The discussion continues in Sect.~\ref{Challenges} with additional
tests to verify independently the results 
summarized in Sect.~\ref{Summary}.

\section{Sagittarius dwarf galaxy: age \& metallicity}
\label{SDG}
The photometric metallicity estimates made for the SDG
thus far depend heavily
on the assumed age. The values for [Fe/H] range from
--0.5 to --1.8 {\it dex}. 
According to Ibata et~al.\ (\cite{Ibata97ea}) 
the mean age (10\to12~Gyr) and metallicity 
(\mbox{[Fe/H]\muspc=\muspc\minus1.5}) adopted is a trade-off, such that
the age conveniently allows for the presence RR Lyrae and carbon stars.
The age range is acceptable for the RR~Lyrae stars, but it is
inconsistent with the evolutionary
calculations for the carbon stars 
mentioned in Sect.~\ref{Introduction}.
\par
Preliminary results, based on HST WFPC2 data,   
reported by 
Mighell et~al.\ (\cite{Mighell98ea}) 
yielded \mbox{[Fe/H]\muspc$\simeq$\muspc\minus0.4}
for a 4 and 6~Gyr 
population\footnote{A (small) revision of the age and 
metallicity is expected in the final results (Mighell 1998)}.
The presence of an even younger population could not be established 
due to the small solid angle covered by the WFPC2.

\section{Isochrones: metallicity and age}
\label{Isochrones}

\subsection{Calibration}
\label{calibration}
The photometric calibration of the AGB-phase of the isochrones 
is a cumbersome process. 
Ng (\cite{Ng97}) gave in his analysis no consideration to the possibility
that the photometric estimate of the metallicity 
might have been too high, which could be due to the uncertainties
of the various transformations applied.
In particular, the photometric calibration of the 
AGB-phase of the isochrones 
in the ESO photometric system (Bouchet et~al.\ \cite{Bouchet91ea}).
The photometric calibration of this phase was not specifically 
based on carbon stars. The dusty atmospheres of carbon stars 
should result in redder colours for a particular metallicity
and {\sc Ng (\cite{Ng97}) likely overestimated the  
metallicity of the ALRW91 C-stars}.
\begin{figure}
\resizebox{\hsize}{!}{\includegraphics{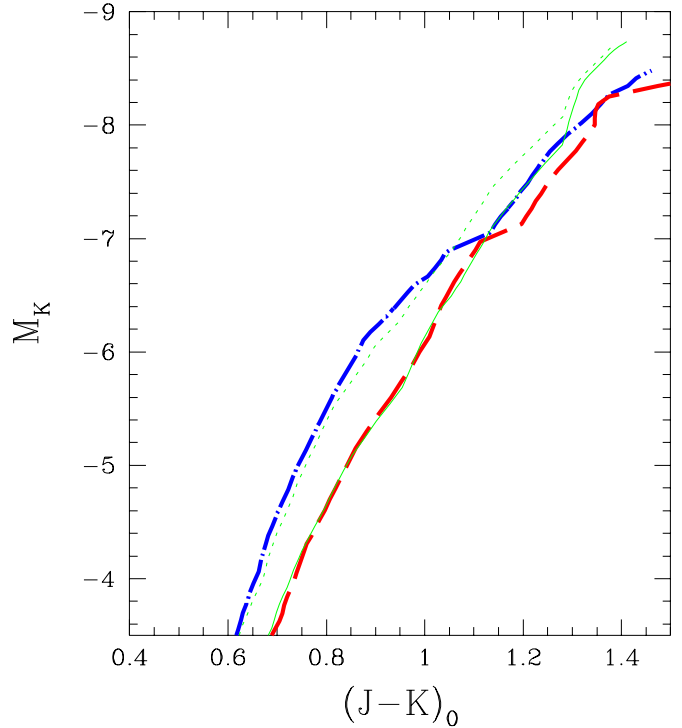}}
\caption{A comparison between isochrones of 1~Gyr with and without dusty
envelopes (carbon grain mixture B; 
Bressan et~al.\ \cite{Bressan98ea}). 
A dotted and a solid line are used for the dust free envelopes 
for \mbox{Z\muspc=\muspc0.004} and \mbox{Z\muspc=\muspc0.008};
for respectively the same chemical compositions
a dot, long dashed and a long dashed line are used for 
the isochrones with dusty envelopes 
}
\label{fig1}
\end{figure}

\subsection{Metallicity from `dusty' isochrones}
\label{metallicity}
Preferably one ought to calibrate correctly the AGB phase for
the carbon stars. Unfortunately, an empirical calibration cannot
be established in the ESO photometric system, 
because not enough data is available. 
An indication of the correction
to be applied to
the metallicity estimated by Ng (\cite{Ng97}) for the ALRW91 C-stars
can be obtained 
from the relative shift between the 
isochrones with and without dusty envelopes computed by 
Bressan et al.\ (\cite{Bressan98ea}; mixture~B with 
amorphous carbon grains) for the SAAO photometric system
(Carter 1990).
The shift is the same for the ESO photometric system.
\par
Figure~\ref{fig1} indicates that 
in the thermally pulsing (TP) AGB phase 
the \mbox{Z\muspc=\muspc0.004} `dusty' isochrone
with a \mbox{$K$-magnitude} 
brighter than \mbox{M$_K$\muspc$\simeq$\muspc\to7\mag0}
is comparable to the \mbox{Z\muspc=\muspc0.008}
non-dusty isochrone.
The isochrones start to diverge from each other 
when the superwind phase sets in at 
\mbox{M$_K$\muspc$\simeq$\muspc\to8\mag2}.
The transition from the E-AGB (early-AGB) to the TP-AGB phase 
is likely smoother than presently modeled for the 
isochrones\footnote{Inclusion of the long-lived subluminous stage
for carbon stars
(Boothroyd et~al.\ \cite{Boothroyd93ea}, Marigo \cite{Marigo98})
would give in the $K$-magnitude range \minus5\mag0
to \minus7\mag0 a still unaccounted, but comparable redward shift of 
the dusty isochrones. 
At fainter luminosities the amount of the 
redward shift of the isochrone, if any,
is unknown. 
An additional complication is that the envelope is chemically enriched
during the TP-AGB phase. The enrichment of the envelope
has also not been included
in the generation of the isochrones.}.
Improvements are however
beyond the scope of
this paper. For our purpose 
Fig.~\ref{fig1} sufficiently demonstrates that
the metallicity of the carbon stars in the TP-AGB phase 
are overestimated with non dusty isochrones.
It is therefore argued that 
{\sc the photometric metallicity estimated by Ng (\cite{Ng97}) 
for the ALRW91 C-stars should 
be revised downward to \mbox{Z\muspc$\simeq$\muspc0.004}}.

\subsection{Age}
\label{age}
Low metallicity carbon stars with an extended carbon envelope
might be confused with stars with less extended envelopes, which 
are either older but have the same metallicity or 
younger/same age and metal-richer.
In addition, the envelope of the carbon star is enriched 
progressively during each thermal pulse.  
As a result the star becomes metal-richer (redder) 
after each thermal pulse for virtually the same age,
i.e. the metallicity enrichment takes precedence over aging. 
The red edge of the carbon stars in the 
(K,J{\minus}K) CMD, see Fig.~2b (Ng \cite{Ng97}), 
is therefore due to metal enriched 
carbon stars with the same age and initial metallicity
as those found at the blue edge. 
This implies that the upper age limit of 1~Gyr given 
by Ng (\cite{Ng97}) for the ALRW91 C-stars is wrong.
\hfill\break
{\sc The age of the ALRW91 C-stars is about 
0.1~Gyr}\footnote{Note that the present results 
suggest that the ALRW91 C-stars and the carbon stars in the Fornax dwarf 
spheroidal galaxy (Stetson et~al.\ \cite{Stetson98ea}) have a
comparable age, but a different metallicity.}.

\section{The velocity dispersion of the carbon stars}
\label{Vdispersion}
The broad velocity dispersion of  
\mbox{$\sigma_{\rm RV}$\muspc=\muspc113\muspc$\pm$\muspc14~km~s$^{-1}$}
(TR91) of the ALRW91 C-stars 
is apparently comparable with a Galactic Bulge dispersion.
A Galactic Bulge membership
leads to the mystery outlined in 
Sect.~\ref{Introduction} and the question arises
if there is an alternative explanation for the velocity dispersion, 
which does not imply a Bulge membership.
In the following sub-sections various alternatives
related to the Sagittarius dwarf galaxy are explored.

\subsection{Are they member of the Sagittarius dwarf galaxy ?}
\label{SDGmember}
The average heliocentric radial velocity of the SDG stars is 
\mbox{$V_R$\muspc=\muspc140\muspc$\pm$\muspc2~km~s$^{-1}$},
which corresponds in galacto-centric coordinates with
\mbox{172~km~s$^{-1}$} and a dispersion of  
\mbox{$\sigma_{\rm RV}$\muspc=\muspc11.4\muspc$\pm$\muspc0.7~km~s$^{-1}$} 
(Ibata et~al.\ \cite{IGI95,Ibata97ea}).
The average radial velocity of the ALRW91 C-stars is
\mbox{$V_R$\muspc=\minus44\muspc$\pm$\muspc20~km~s$^{-1}$}.
\hfill\break
It is immediately evident from both the average velocities and 
their dispersion that the majority or even all of the ALRW91 C-stars 
cannot be member of the SDG.
\par
Numerical calculation (see
Edelsohn{\muspc\&\muspc}Elmegreen \cite{EE97},
Johnston et~al.\ \cite{Johnston95ea} and 
Vel\'azquez{\muspc\&\muspc}White \cite{VelazquezWhite95})
indicated that the velocity dispersion of the SDG stars
does not change significantly on approaching and after crossing of the 
Galactic plane. 
Johnston et~al.\ also demonstrated for
moving groups with stars stripped from the dwarf galaxy,
that their radial velocities change but that they do 
maintain a small velocity dispersion.
\hfill\break
{\sc The ALRW91 C-stars cannot be member of the SDG, nor can they be a tidally 
stripped moving group.}

\subsection{Are they formed during crossing the disc? }
\label{crossing}

\subsubsection{In the Galactic anti-centre direction ?}
\label{anti-centre}
If the SDG is moving towards the galactic mid-plane 
(Edelsohn{\muspc\&\muspc}Elmegreen \cite{EE97}
and references cited therein)
then the most recent crossing occurred in the direction of the 
galactic anti-centre. 
At the present position this should have resulted in a
small velocity dispersion similar to the one of the SDG stars.
The broad velocity distribution 
of the C-stars does not support the possibility that
these stars have been dragged along the SDG orbit from the anti-centre
to its present position. 
\hfill\break
{\sc The ALRW91 C-stars were not formed in the Galactic anti-centre
direction}.
\begin{figure*}
\resizebox{10cm}{!}{\includegraphics{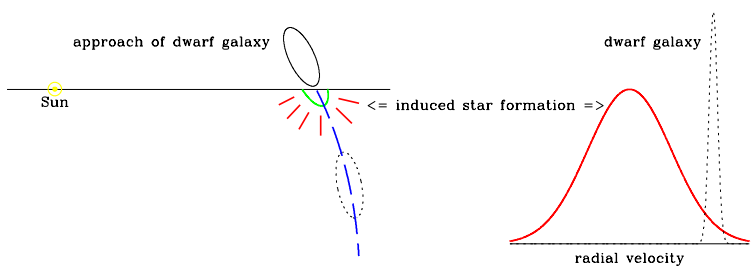}}
\hfill
\parbox[b]{75mm}{%
\caption{Schematic view of the broad, bulge like, radial 
velocity distribution for the material from an induced star 
formation event, caused by the crossing from a dwarf galaxy 
through the Galactic mid-plane.
Note that the majority of the new material will not be moving 
along the dwarf galaxy orbit. The dotted ellipse and the 
dotted line indicate respectively the 
current position and radial velocity (not on scale) for the dwarf galaxy
\label{InducedStarFormation}}}
\end{figure*}

\subsubsection{On the recent approach towards the disc ?}
\label{approach}
Hydrodynamical calculations of 
the interaction of the SDG with a gaseous H~I disc 
by Ibata\&Razoumov (\cite{IbataRazoumov98})
indicate that on approach of the SDG part of the H~I layer 
is first pulled out of the disc 
due to the attractive influence of the dwarf galaxy.
The temperature and density of the gas is at this stage not favourable
to turn part of the gas into stars. This will change when the SDG gets 
nearer to the midplane and part of the material is
compressed to densities high enough to sustain star formation.
\hfill\break
The SDG is located
at about 6~kpc out of the Galactic midplane.
This is too far away to invoke star formation. 
\hfill\break
{\sc The ALRW91 C-stars are not formed on approach of
the Galactic midplane}.

\subsubsection{On a recent crossing ?}
\label{crossed}
Star formation could be invoked
from the material residing in either the SDG and/or near 
the impact spot in the Galactic disc,
when the SDG is nearby the galactic plane and crosses it.
Stars formed from SDG material are SDG members and 
can be ignored in this discussion
(see Sect.~\ref{SDGmember}).
Near the Galactic plane the SDG starts to compress the material 
that it pulled out from the disc earlier on its approach 
(see Sect.~\ref{approach}). Star formation is mainly triggered
during and after the crossing of the Galactic 
plane\footnote{According to Ibata{\muspc\&\muspc}Razoumov 
(\cite{IbataRazoumov98})
there is substantial heating only in disc shocks and tidal tails
which follow the SDG into the halo.
They did not consider in their calculations 
star formation in the disc shocks and the tidal tails.
However, their number density-temperature diagram (their Fig.~4)
suggests that, following generally adopted star formation
scenarios (see for example Carraro et~al.\ \cite{CLC98} and 
references cited therein),
part of their gaseous material (\mbox{log~T\muspc$\simeq$\muspc2\to4})
should have been converted 
into stars. Moreover, their Fig.~2b indicate that
just after the collision a combination of gaseous and stellar
material is scattered in all directions. 
A detailed analysis of the radial velocities of the 
scattered material should confirm the 
feasibility of the scenario outlined above.
}.
\hfill\break
The SDG pushes material out of the Galactic disc. 
A small fraction of the newly 
formed stars is dragged along the orbit, while the majority
of the material together with some young stars is moving away from the SDG.
Part of these young stars are moving with different velocities
towards us while another part of the stars are moving away from us.
As a consequence, the distribution
of radial velocities is considerably broader than expected 
from a SDG membership alone. The resulting distribution
can even mimic a Bulge-like velocity dispersion.
\hfill\break
Figure~\ref{InducedStarFormation} gives a schematic view 
of this process.
If the motion of the dwarf galaxy is perpendicular 
to the galactic plane then the resulting radial velocity distribution 
after crossing the galactic plane will be symmetric 
around zero. The situation sketched in 
Fig.~\ref{InducedStarFormation} implies that more stars
will be found moving towards us, thus giving a negative value
for the average radial velocity.
\hfill\break
{\sc A recent star formation event, induced by the crossing through 
the galactic plane of the SDG, can account for both  
the average radial velocity and the velocity dispersion 
of the ALRW91 C-stars.}

\section{Motion}
\label{Motion}

\subsection{Which direction ?}
\label{direction}
The models of the SDG orbit indicate 
(see Fig.~11 Ibata et~al.\ \cite{Ibata97ea} and references cited therein), 
together with the {\em preliminary}\/ proper motion reported by 
Irwin et~al.\ (\cite{Irwin96ea}), that
the SDG is moving towards the galactic mid-plane.
On the other hand,
the ALRW91 C-stars appear to form the evidence that the 
SDG already crossed the galactic mid-plane. Furthermore,
the Ibata et al.\ orbit at low galactic latitudes
appears to be inconsistent with the position 
obtained from RR~Lyrae stars,
see Fig.~\ref{fig4} and Sect.~\ref{where_crossing}.
A study of the Galactic globular cluster Palomar~5 
(Scholz et~al.\ \cite{Scholz98ea}) indicates further that the cluster 
is moving in the opposite 
direction\footnote{The work of Scholz et~al.\ demonstrates
how the direction of space motion in Galacto-centric terms 
changed with respect to the results from 
the {\it preliminary}
proper motion reported by Schweitzer et~al.\ (\cite{Schweitzer93ea}).} 
with respect to the
Ibata et~al.\ orbital motion of the SDG. 
\hfill\break
Instead of looking for alternative explanations for the 
apparent contradictions, one should consider 
an independent determination of the proper motion
with a zeropoint tied to one or more distant galaxies.
\par
Recognizing that the present contradiction about the direction of 
motion of the SDG will not be solved until {\it definite}
proper motions are available,
and considering that reasonable grounds are given
to use a lower weight for the {\it preliminary}\/ 
value of the proper motion
reported for the SDG,
I assume for the remaining part of the paper that 
{\sc the SDG is not moving towards the Galactic midplane,
but crossed it recently.}

\subsection{Where did the last crossing occurred ?}
\label{where_crossing}
The impact position at the galactic mid-plane 
is obtained from an unweighted least-squares
fit through the distances determined for the SDG from RR Lyrae stars
(Alcock et~al.\ \cite{MACHO97}, Alard \cite{Alard96},
NS97, and Mateo et~al.\ 
\cite{Mateo95ea}\&\cite{Mateo96ea})\footnote{Alcock 
et~al.\ estimated the distance by scaling the 2\mag2
difference between the Bulge and SDG, assuming implicitly
\mbox{$R_\odot$\muspc=\muspc8~kpc}.
The remaining distances were determined with an adopted RR Lyrae 
luminosity: Alard, NS97, and Mateo et~al.\ (\cite{Mateo95ea})
adopted \mbox{$M_V$\muspc=\muspc0\mag6}, while 
Mateo et~al.\ (\cite{Mateo96ea}) used \mbox{$M_V$\muspc=\muspc0\mag8}.
The latter gives an inhomogeneity in the distances. 
There is no reason to assume
that the luminosity of the SDG RR~Lyrae stars is different at various
galactic latitudes. The distance of 27.3~kpc obtained by 
Mateo et~al.\ (\cite{Mateo96ea})
was therefore adjusted to \mbox{24.9\muspc$\pm$\muspc1.8~kpc}.
}
as a function of the galactic latitude.
\par
\begin{figure}
\resizebox{\hsize}{!}{\includegraphics{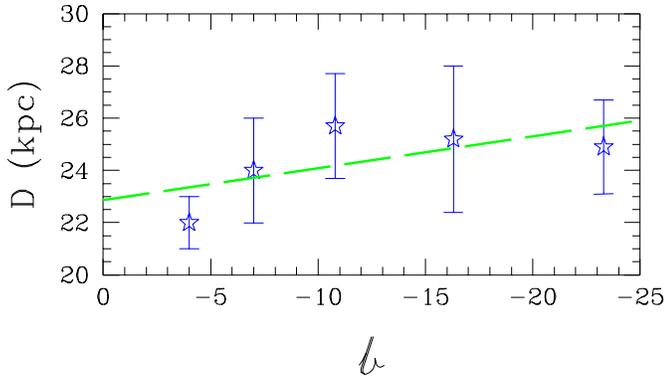}}
\caption{The distance from RR Lyrae stars
to the Sagittarius dwarf galaxy  
as a function of galactic latitude.
The assumed RR~Lyrae luminosity is \mbox{$M_V$\muspc=\muspc0\mag6}. 
The long dashed line is determined from an unweighted least-squares 
fit and yields:\ {\it D}\/(kpc)\muspc=\muspc$22.83-0.12b$.
}
\label{fig3}
\end{figure}
\begin{figure}
\resizebox{\hsize}{!}{\includegraphics{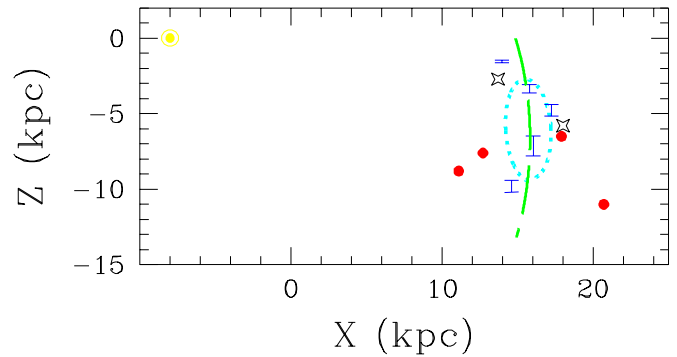}}
\caption{The course (long dashed line) of the Sagittarius dwarf galaxy 
(dotted ellipse) projected on
the $x-z$ plane. The Sun ($\odot$) is positioned at (\minus8,0).
The error bars indicate the average location of the RR~Lyrae stars
displayed in Fig.~\ref{fig3}. Skeletal crosses are used for two
carbon stars (one Mira variable, see Whitelock \cite{Whitelock98};
one semiregular variable, see NS97
and Schultheis et~al.\ \cite{Schultheis98ea}), 
and filled dots are used for 
the four globular clusters associated with Sagittarius dwarf galaxy
(Da Costa{\muspc\&\muspc}Armandroff
\cite{DaCostaArmandroff95}, Ibata et~al.\ \cite{Ibata97ea})
}
\label{fig4}
\end{figure}

Figure~\ref{fig3} displays the distances from the RR~Lyrae stars 
as a function of galactic latitude.
An unweighted least-squares fit 
gives {\it D}\/(kpc)\muspc=\muspc$22.83-0.12b$.
One thus obtains
22.8~kpc for the distance towards the disc impact position of the SDG.
Adopting a Solar galacto-centric distance of 
\mbox{R$_0$\muspc=\muspc8.0~kpc} 
(Paczy\'nski{\muspc\&\muspc}Stanek \cite{PS98}, 
Wesselink \cite{Wesselink87})
implies that {\sc the crossing of the SDG through the Galactic plane
occurred behind the Galactic Bulge at 14.8~kpc from
the Galactic centre}.

\subsection{Are the carbon stars related to the Galactic disc ?}
\label{in_disc}
A consistency check is made for the suggestion that the SDG already 
crossed the Galactic plane and triggered a star formation event in the 
Galactic disc (Sect.~\ref{crossed}). 
Such an event would imply that the disc metallicity
at impact position has to be comparable with the photometric
determination of the metallicity of the ALRW91 C-stars.
To determine the disc metallicity at impact position one 
has to take the radial dependence of [Fe/H] into account.
Carraro et~al.\ (\cite{CNP98}) determined from open clusters 
the radial metallicity gradient in different age ranges. 
\hfill\break
For the clusters younger than 2~Gyr a present day gradient of
\mbox{\minus0.063\muspc$\pm$\muspc0.013} {\it dex}~kpc$^{-1}$ was obtained.
For all clusters in the sample the average gradient is
\mbox{\minus0.085\muspc$\pm$\muspc0.008} {\it dex}~kpc$^{-1}$.
The normalization at the Solar position is respectively
\mbox{\minus0.18\muspc$\pm$\muspc0.12} {\it dex}\/ and 
\mbox{\minus0.15\muspc$\pm$\muspc0.08} {\it dex}.
The disc metallicity at the crossing position thus obtained is
\mbox{[Fe/H]\muspc=\muspc\to0.61\muspc$\pm$\muspc0.20}
(\mbox{Z\muspc$\simeq$\muspc0.0050})
or \mbox{[Fe/H]\muspc=\muspc\to0.73\muspc$\pm$\muspc0.10} 
(\mbox{Z\muspc$\simeq$\muspc0.0035})
with respectively the present day and the
average radial metallicity gradient. 
\par
Irrespective of the present day or average metallicity 
gradient it is estimated that {\sc the disc metallicity at the 
crossing position is 
\mbox{Z\muspc$\simeq$\muspc0.0045\muspc$\pm$\muspc0.0010}.
Within the uncertainties the Galactic disc metallicity
at the impact position is the same as metallicity of the ALRW91 C-stars}
(Sect.~\ref{metallicity}).

\subsection{Are the carbon stars near Sagittarius dwarf galaxy ?}
\label{where_are_they}
Another point to examine is how far the ALRW91 C-stars
have moved from the impact since the crossing through the Galactic plane
of the SDG. The C-stars were accelerated out of the Galactic midplane
to their present average radial velocity of \mbox{\minus44~km~s$^{-1}$}.
In 0.1~Gyr they could have traveled about 4.4~kpc towards us.
The actual distance traveled is considerably less, 
say \mbox{$\sim$\muspc2.2\to3.1~kpc}, because
the C-stars had a considerable smaller velocity in the past. 
The traveled distance is within the \mbox{10\%\to15\%}
uncertainty of the 22.8~kpc distance to the impact position.
\par
The velocity dispersion of the C-stars provides further an indication
about the average separation. Taking into
account that the dispersion is smaller,
due to a combination of turbulence and a lower velocities in the past,
the separation is estimated to \mbox{$\sim$5.6\to8.0~kpc}.
\hfill\break
This can be compared with
the distance between 2 of the 4~globular clusters associated 
with SDG is about 10~kpc,
i.e. Terzan 8 at 21.1~kpc and Arp~2 at 31.0~kpc 
(Da Costa{\muspc\&\muspc}Armandroff \cite{DaCostaArmandroff95}).
\hfill\break
The distance between the two carbon stars with well determined
periods\footnote{Note that up to date no carbon semiregulars or Miras 
with well determined periods and luminosities have been reported,
for which the period-luminosity relation unambiguously places 
them at a Galactic Bulge distance.}
is on the other hand about 5 kpc, i.e. 
21.9~kpc for a carbon semiregular variable (NS97, 
Schultheis et~al.\ \cite{Schultheis98ea} 
and 26.7~kpc for a carbon Mira (Whitelock \cite{Whitelock98}).
The separation between the ALRW91 C-stars is from the above
consideration within acceptable limits.
\hfill\break 
{\sc The ALRW91 C-stars are 
still at a distance related with the SDG}.

\subsection{ALRW91 C-stars = SDG C-stars ?}
\label{ALRWeqSDG}
Ibata et~al.\ (\cite{IGI95}) identified four new carbon stars
in a field skimming over the Galactic Bulge.
The radial velocities of the stars confirmed their membership 
to the SDG. Their average \mbox{(J{\minus}K)$_{0,saao}$} colour is 1\mag40.
The average \mbox{(J{\minus}K)$_{0,eso}$}
colour for the ALRW91 C-stars is 0\mag85.
The colour difference indicates that there should be  
marked differences between the ALRW91 and the SDG 
C-stars\footnote{Note that the semiregular carbon variable, mentioned
in Sect.~\ref{where_are_they}, belongs to the `blue' group,
while the carbon Mira is clearly related to the `red' group
of carbon stars.}, 
because comparable colours are expected if they 
originated from the same star formation event.
\par
The difference between the two groups is an indication for  
differences in age and metallicity.
The SDG C-stars should belong to a stellar population  
with an age comparable or younger than $\sim$4~Gyr.
Only the youngest stellar population identified thus far for the SDG
matches this constraint: the population  
has an age of 4~Gyr and a metallicity around 
\mbox{Z\muspc$\simeq$\muspc0.008} 
(Mighell et~al.\ \cite{Mighell98ea}). 
This is significantly different from the metallicity and age 
obtained for
the ALRW91 C-stars: \mbox{Z\muspc=\muspc0.004} and 0.1~Gyr
(see Sects.~\ref{metallicity} \& \ref{age}). 
\hfill\break 
{\sc The ALRW91 C-stars $\ne$ the SDG C-stars: the SDG C-stars 
are older and metal-richer}.

\section{Challenges}
\label{Challenges}

\subsection{More carbon stars wanted}
It will become very important to increase the 
sample size of the C-stars related to the SDG. 
Ng (1997) argues that still a considerable number of C-stars 
could be found in the databases with long period variable stars 
from the various micro-lensing projects 
(see the contributions described in Ferlet et~al.\ 1997).
\par
The distances of these stars can be obtained from
their periods and K-band luminosities.
The velocity dispersion of a significant number of
carbon Miras \& semiregular variables located 
at a distance comparable to the SDG 
will provide an independent verification of the hypothesis
that the ALRW91 C-stars are the result of an induced star 
formation event (see Sect~\ref{crossed}).
\hfill\break
Note that special care should be taken to avoid mixing the 
older, metal-richer SDG carbon stars with the essentially 
younger ALRW91-like C-stars\footnote{A very 
young age of \mbox{$\sim$\muspc0.1~Gyr} for the ALRW91 C-stars 
would further anticipate in the microlensing databases the presence 
of a considerable number of Cepheids.
In particular the anomalous faint Cepheids are 
interesting, due to a possible analogy 
with the former 2\mag5 mystery of the ALRW91 C-stars.  
} 

\subsection{Comparison with theoretical models}
The ALRW91 \& the SDG C-stars offer a great opportunity to
study along the same line of sight 
two distinct populations of carbon stars (Sect.~\ref{ALRWeqSDG}).
The distribution of both the ALRW91 and SDG C-stars ought 
to be compared with the 
theoretical models from Marigo (1998) using for each group
respectively the SMC and LMC metallicity. 
Moreover, the difference in metallicity and the 
similarity in age between the ALRW91 C-stars and those
from the Fornax dwarf galaxy (Stetson et~al.\ \cite{Stetson98ea}) 
provides 
furthermore the possibility to tune the theoretical models to even
lower metallicities.
\hfill\break
Such a 
comparison would provide
an independent verification of the age of each group of C-stars 
and support the assertion that the
ALRW91 C-stars did form from material originating from
our Galactic disc.
\hfill\break
It is not clear what kind of event lead to the formation of the SDG C-stars
\mbox{$\sim$\muspc4~Gyr} ago.
It is possibly a time-stamp of the encounter with the LMC which 
deflected the SDG in a closer orbit around our Galaxy (Zhao 1998).
This encounter should have left in the LMC traces of a 
population of carbon stars with a comparable age.

\subsection{Radial velocities of young stars}
In Sect.~\ref{crossed} it is mentioned that the 
hydrodynamical calculations of the interaction between 
the SDG and the Galactic H~I disc (Ibata\mbox{\muspc\&\muspc}Razoumov
\cite{IbataRazoumov98}) provide an indication that, after a collision
of the SDG with the Galactic plane, gaseous and stellar 
material will be scattered around in all directions.
The scattering results in a small number of young stars
in the CMDs from Marconi et~al.\ 
(\cite{Marconi98ea}, see also Fig.~3 Ng \cite{Ng97}).
A comparison of the velocity dispersion between the
ALRW91 C-stars and the young stars found at a SDG related 
distance should provide an independent verification  
of the induced star formation scenario suggested in
this paper.

\subsection{Chemo-dynamical formation models}
The models thus far consider only the gravitational interaction between
the Galaxy and the SDG (Ibata\mbox{\muspc\&\muspc}Lewis \cite{IbataLewis98},
Ibata\mbox{\muspc\&\muspc}Razoumov \cite{IbataRazoumov98}, and
Nair\mbox{\muspc\&\muspc}Miralda-Escud\'e \cite{Nair98}), 
but do not take into account
an encounter, induced, star formation event.
Improved models should therefore be employed to trace the stars
from such an event.
\par
The SDG offers the opportunity to
study in great detail encounter induced star formation events. 
Chemo-dynamical models (Carraro et~al.\ 1998a, Gerritsen 1997)
should be explored, which take into account the induced star formation 
when a SDG like object crosses our Galactic disc, 
to determine when star formation occurs,
when carbon stars will emerge, and 
to determine to which extent the newly formed stars can be dragged 
along the orbit. 
\par
As an aside, induced star formation due to the crossing of a 
dwarf spheroidal galaxy through the galactic mid-plane might be 
one of the processes responsible for the elusive nature of the so-called
halo carbon stars (see Groenewegen et~al.\ \cite{Groenewegen97ea}
and references cited therein).
Hydrodynamical calculations of the interaction between 
the SDG and the Galactic H~I disc (Ibata\mbox{\muspc\&\muspc}Razoumov
\cite{IbataRazoumov98}) indicate that after a collision
material from the H~I disc is
pulled/pushed out of the plane to \mbox{$\sim$\muspc10~kpc} heights.
i.e. comparable to the heights of the halo carbon stars.
This material is partly made up out of stars (see Sect.~\ref{crossed})
results in a prominent signature and remains clearly 
visible for at least half an orbit.

\section{Summary}
\label{Summary}
The results obtained from the analysis by Ng (\cite{Ng97})
of the hypothesis that the ALRW91 C-stars are located at a SDG related
distance have been improved.
\hfill\break
{$\bullet$} The metallicity of the ALRW91 C-stars is 
\mbox{Z\muspc$\simeq$\muspc0.004}.
\hfill\break
{$\bullet$} The age of the ALRW91 C-stars is $\sim$0.1 Gyr. 
\hfill\break
In this paper the average velocity and the velocity dispersion of the 
ALRW91 C-stars were taken into consideration. 
They added the conditions that the stars
\hfill\break
{$\circ$} cannot be member of the SDG;
\hfill\break
{$\circ$} are not a tidally stripped moving group;
\hfill\break
{$\circ$} were not formed on the approach of the SDG 
towards the Galactic midplane;
\hfill\break
{$\bullet$} have to be formed during a recent crossing of the SDG through
the Galactic plane at about 14.8~kpc behind the Galactic Bulge. 
\par
In addition, a comparison of the near-infrared photometry 
between the SDG \& the ALRW91
C-stars indicated that they do not originate from the same 
star formation event: the SDG C-stars are older and metal-richer
than the ALRW91 C-stars.
\par
The condition that the SDG must have crossed the Galactic plane
will remain a point of dispute until {\it definite}\/
proper motions are obtained with the HST 
or GAIA (Global Astrometric Interferometer for Astrophysics).

\begin{acknowledgements}{C.~Chiosi and an anonymous referee are
thanked for their critical comments 
on this paper.
A.~Bressan provided not only additional food for thought 
but also the dusty isochrones with 
a carbon grain mixture for Fig.~\ref{fig1}. G.~Carraro,
L.~Girardi, M.A.T.~Groenewegen and P.~Marigo 
are acknowledged for providing additional suggestions.
The research of Y.K.~Ng is supported by 
TMR grant ERBFMRX-CT96-0086 from the European Community.}
\end{acknowledgements}


\begin{thebibliography}{}
\bibitem [1996]{Alard96}
Alard C., 1996, ApJ 458, L17
\bibitem [1997]{MACHO97}
Alcock C., Allsman R.A., Alves D.R., et~al., 1997, ApJ 474, 217
\bibitem [1994] {Azzopardi94}
Azzopardi M., 1994, in proceedings third ESO/CTIO workshop 
{\it `The Local Group: comparative and global properties'},
A. Layden, R.C. Smith and J. Storm (eds.), 129
\bibitem [1985] {ALR85}
Azzopardi M., Lequeux J., Rebeirot E., 1985, A\&A 145, L4
\bibitem [1988] {ALR88}
Azzopardi M., Lequeux J., Rebeirot E., 1988, A\&A 202, L27 
\bibitem [1991] {ALRW91}
Azzopardi M., Lequeux J., Rebeirot E., Westerlund B.E., 1991, A\&AS 88, 265
(ALRW91)
\bibitem [1993] {Boothroyd93ea}
Boothroyd A.I., Sackmann I.-J., Ahern S.C., 1993, ApJ 416, 762 
\bibitem [1991] {Bouchet91ea}
Bouchet P., Manfroid J., Schmider F.X., 1991, A\&AS 91, 409
\bibitem [1998] {Bressan98ea}
Bressan A., Granato G.L., Silva L., 1998, A\&A 332, 135

\bibitem [1998a]{CLC98}
Carraro G., Lia C., Chiosi C., 1998a, MNRAS 297, 1021
\bibitem [1998b]{CNP98}
Carraro G., Ng Y.K., Portinari L., 1998b, MNRAS 296, 1045 
\bibitem [1990] {Carter90}
Carter B.S., 1990, MNRAS 242, 1
\bibitem [1992]{CBB92}
Chiosi C., Bertelli G., Bressan A., 1992, ARA\&A 30, 235
\bibitem[1995]{DaCostaArmandroff95}
Da Costa G.S., Armandroff T.E., 1995, AJ 109, 2533

\bibitem [1997]{EE97}
Edelsohn D.J., Elmegreen B.G., 1997, MNRAS 290, 7
\bibitem [1997]{Ferlet97ea}
Ferlet R., Maillard J.-P., B. Raban, 1997, 12th IAP Colloquium 
{\it `Variable stars and the astrophysical returns from microlensing surveys'},
Paris (France), 8\to12 July 1996,
Editions Fronti\'eres
\bibitem [1997]{Gerritsen97}
Gerritsen J.P.E., 1997, Ph.D. thesis, Groningen University, the
Netherlands
\bibitem [1993]{GdJ93}
Groenewegen M.A.T., de Jong T., 1993, A\&A 267, 410
\bibitem [1995]{Groenewegen95ea}
Groenewegen M.A.T., van den Hoek L.B., de Jong T., 1995, A\&A 293, 381
\bibitem [1997]{Groenewegen97ea}
Groenewegen M.A.T., Oudmaijer R.D., Ludwig H.G., 1997, MNRAS 292, 686

\bibitem[1998]{IbataLewis98}
Ibata R.A., Lewis G.F., 1998, ApJ 500, 575 
\bibitem[1998]{IbataRazoumov98}
Ibata R.A., Razoumov A.O., 1998, A\&A 336, 130
\bibitem[1994]{IGI94}
Ibata R.A., Gilmore G., Irwin M.J., 1994, Nature 370, 194
\bibitem[1995]{IGI95}
Ibata R.A., Gilmore G., Irwin M.J., 1995, MNRAS 277, 781
\bibitem[1997]{Ibata97ea}
Ibata R.A., Wyse R.F.G., Gilmore G., Irwin M.J., Suntzeff N.B., 
1997, AJ 113, 635

\bibitem [1996]{Irwin96ea}
Irwin M., Ibata R., Gilmore G., et~al.\, 1996, in proceedings 
{\it `Formation of
the galactic halo ... inside and out'}, 
Tucson, Arizona (USA), 9\to11 October 1995,
ASP conference Series 92,
H. Morrison and A. Sarajedini (eds.), 84
\bibitem [1995]{Johnston95ea}
Johnston K.V., Spergel D.N., Hernquist L., 1995, ApJ 451, 598
\bibitem [1989]{Lattanzio89}
Lattanzio J.C., 1989, ApJ 344, L25
\bibitem [1990]{Lequeux90}
Lequeux J., 1990, in proceedings {\it`From Miras to planetary nebulae:
which path for stellar evolution?'}, Montpellier (France), 4\to7 September
1989, M.-O.~Mennessier and A. Omont (eds.), Editions Fronti\'eres, 273

\bibitem[1998]{Marconi98ea}
Marconi G., Buonanno R., Castellani M., et al. 1998, A\&A 330, 453 
\bibitem[1998]{Marigo98}
Marigo P., 1998, Ph.D. thesis, Padova University, Italy
\bibitem[1996a]{Marigo96aea}
Marigo P., Bressan A., Chiosi C., 1996a, A\&A 313, 545
\bibitem[1996b]{Marigo96bea}
Marigo P., Girardi L., Chiosi C., 1996b, A\&A 316, L1

\bibitem[1995]{Mateo95ea}
Mateo M., Kubiak M., Szyma\'nski M., et al., 1995, AJ 110, 1141
\bibitem[1996]{Mateo96ea}
Mateo M., Mirabel N., Udalski A., et al., 1996, ApJ 458, L13
\bibitem[1998]{Mighell98}
Mighell K., 1998, private communication
\bibitem[1998]{Mighell98ea}
Mighell K., Armandroff T., Sarajedini A., et~al., 1998,
in proceedings IAU symposium 186 
{\it `Galaxy interaction at low and high redshift'}, 
Kyoto (Japan), 26\to30 August 1997,
ed.~D.B.\ Sanders (Kluwer: Dordrecht), {\it in press}
\bibitem[1998]{Nair98}
Nair V., Miralda-Escud\'e J., 1998, ApJ {\it submitted}
({\tt astro-ph/9805278})
\bibitem[1997]{Ng97}
Ng Y.K., 1997, A\&A 328, 211
\bibitem[1998]{Ng98}
Ng Y.K., 1998,
in proceedings IAU symposium 186 
{\it `Galaxy interaction at low and high redshift'}, 
Kyoto (Japan), 26\to30 August 1997,
ed.~D.B.\ Sanders (Kluwer: Dordrecht),{\it in press}
\bibitem [1997]{NS97}
Ng Y.K., Schultheis M., 1997, A\&AS 123, 115 (NS97)

\bibitem [1998]{PS98}
Paczy\'nski B., Stanek K.Z., 1998, ApJ 494, L219  
\bibitem [1998]{Scholz98ea}
Scholz R.-D., Irwin M,, Odenkirchen M., Meusinger H., 1998, A\&A 333, 531
\bibitem [1998]{Schultheis98ea}
Schultheis M., Ng Y.K., Hron J., Kerschbaum F., 
1998, A\&A {\it accepted}
\bibitem[1993]{Schweitzer93ea}
Schweitzer A.E., Cudworth K., Majewski S., 1993, ASP Conf. Ser. 48, 113
\bibitem [1998]{Stetson98ea}
Stetson P.B., Hesser J.E., Schmecker-Hane T.A., 1998, PASP 110, 533
\bibitem [1991]{TR91}
Tyson N., Rich R.M., 1991, ApJ 367, 547 (TR91)
\bibitem[1995]{VelazquezWhite95}
Vel\'azquez H., White S.D.M., 1995, MNRAS 275, L23
\hyphenation{Nij-me-gen}
\bibitem [1987] {Wesselink87}
Wesselink Th.J.H., 1987, Ph.D. thesis, Catholic University
Nijmegen, the Netherlands
\bibitem [1991] {Westerlund91ea}
Westerlund B.E., Lequeux J., Azzopardi M., Rebeirot E., 1991, A\&A 244, 367
\bibitem [1995] {Westerund95ea}
Westerlund B.E., Azzopardi M., Breysacher J., Rebeirot E., 1995, A\&A 303, 107
\bibitem[1998]{Whitelock98}
Whitelock P.A., 1998, to appear in {\it `Pulsating Stars - Recent
Developments in Theory and Observation'}, 
M. Takeuti \& D. Sasselov (eds.), Universal Academic Press,
Tokyo, {\it in press}
\bibitem [1998]{Zhao98}
Zhao H.S., 1998, ApJ 500, L149
\end{thebibliography}
\end{document}